\documentclass[iop,twocolumn]{emulateapj}
\bibpunct[; ]{(}{)}{;}{a}{}{;}
\usepackage{apjfonts}

\usepackage{graphicx}
\usepackage{natbib}

\newcommand{\Msolar}{M${_\odot}$\,}

\shorttitle{Star formation in 30 Doradus}
\shortauthors{De Marchi et al.}

\received{2 May 2011}
\accepted{13 June 2011}

\slugcomment{}

\begin{document}

\title{Star formation in 30 Doradus\,\altaffilmark{*}}


\author{Guido De Marchi,\altaffilmark{1} 
Francesco Paresce,\altaffilmark{2}
Nino Panagia,\altaffilmark{3,4,5}
Giacomo Beccari,\altaffilmark{6} 
Loredana Spezzi,\altaffilmark{1} \\
Marco Sirianni,\altaffilmark{1}
Morten Andersen,\altaffilmark{1}
Max Mutchler,\altaffilmark{3}
Bruce Balick,\altaffilmark{7}
Michael A. Dopita,\altaffilmark{8} 
Jay A. Frogel,\altaffilmark{9} \\ 
Bradley C. Whitmore,\altaffilmark{3}
Howard Bond,\altaffilmark{3}
Daniela Calzetti,\altaffilmark{10}
C. Marcella Carollo,\altaffilmark{11}
Michael J. Disney,\altaffilmark{12} \\
Donald N. B. Hall,\altaffilmark{13}
Jon A. Holtzman,\altaffilmark{14}
Randy A. Kimble,\altaffilmark{15}
Patrick J. McCarthy,\altaffilmark{16}
Robert W. O'Connell,\altaffilmark{17} \\
Abhijit Saha,\altaffilmark{18}
Joseph I. Silk,\altaffilmark{19}
John T. Trauger,\altaffilmark{20}
Alistair R. Walker,\altaffilmark{21}
Rogier A. Windhorst\altaffilmark{22} \\
and Erick T. Young\altaffilmark{23} }

\altaffiltext{1}{European Space Agency, Space Science Department, Keplerlaan
1, 2200 AG Noordwijk, Netherlands; gdemarchi@rssd.esa.int}
\altaffiltext{2}{Istituto di Astrofisica Spaziale e Fisica Cosmica, Via
Gobetti 101, 40129 Bologna, Italy}
\altaffiltext{3}{Space Telescope Science Institute, 3700 San Martin
Drive, Baltimore MD 21218, USA}
\altaffiltext{4}{INAF--CT, Osservatorio Astrofisico di Catania, Via S. Sofia 
78, 95123 Catania, Italy}
\altaffiltext{5}{Supernova Limited, OYV \#131, Northsound Rd., Virgin Gorda, 
British Virgin Islands}
\altaffiltext{6}{European Southern Observatory, Karl--Schwarzschild-Str.
2, 85748 Garching, Germany}
\altaffiltext{7}{Department of Astronomy, University of Washington,
Seattle, WA 98195--1580, USA}
\altaffiltext{8}{Research School of Astronomy \& Astrophysics, The
Australian  National University, ACT 2611, Australia}
\altaffiltext{9}{Galaxies Unlimited, 8726 Hickory Bend Trail, Potomac,
MD 20854, USA}
\altaffiltext{11}{Department of Astronomy, University of Massachusetts, 
Amherst, MA 01003, USA}
\altaffiltext{12}{Department of Physics, ETH--Zurich, Zurich, 8093, 
Switzerland}
\altaffiltext{10}{School of Physics and Astronomy, Cardiff University, 
Cardiff CF24 3AA, United Kingdom}

\begin{abstract}  

Using observations obtained with the Wide Field Camera 3 (WFC3) on board
the  {\it Hubble Space Telescope} (HST), we have studied the properties 
of the stellar populations in the central regions of 30\,Dor, in the
Large Magellanic Cloud. The observations clearly reveal the presence of
considerable differential extinction across the field. We characterise
and quantify this effect using young massive main sequence stars to
derive a statistical reddening correction for most objects in the field.
We then search for pre-main sequence (PMS) stars by looking for objects
with a strong ($> 4\,\sigma$) H$\alpha$ excess emission and find about
1150 of them over the entire field. Comparison of their location in
the Hertzsprung--Russell diagram with theoretical PMS evolutionary
tracks for the appropriate metallicity reveals that about one third of
these objects are younger than $\sim 4$\,Myr, compatible with the age of
the massive stars in the central ionising cluster R\,136, whereas the
rest have ages up to $\sim 30$\,Myr, with a median age of $\sim 12$\,Myr.
This indicates that star formation has proceeded over an extended period
of time, although we cannot discriminate between an extended episode and
a series of short and frequent bursts that are not resolved in time.
While the younger PMS population preferentially occupies the central
regions of the cluster, older PMS objects are more uniformly distributed
across the field and are remarkably few at the very centre of the
cluster. We attribute this latter effect to photoevaporation of the
older circumstellar discs caused by the massive ionising members of
R\,136. 

\end{abstract}


\section{Introduction}

Many years of effort have been spent on trying to understand the
structure and evolution of the super star cluster NGC\,2070, at the
centre of the 30\,Doradus nebula in the Large Magellanic Cloud
(otherwise known as the Tarantula nebula) since its discovery as such by
Lacaille in 1751. In this case, the prize is clearly worth the effort as
this object may hold the key to answers to many fundamental questions
regarding the physics of the formation of stars, stellar clusters and
galaxies and their role in the cosmos (e.g. Brandl 2005). Thanks also to
this effort, we now know much about the brightest stars more massive
than $\sim 2-3$\,\Msolar that can be more easily detected and resolved.
\footnotetext{Institute for Astronomy, University of Hawaii, Honolulu,
HI 96822, USA}\footnotetext{Department of Astronomy, New Mexico State
University,  Las Cruces, NM 88003, USA} \footnotetext{NASA-Goddard Space
Flight Center, Greenbelt, MD 20771,  USA} \footnotetext{Observatories of
the Carnegie Institution of Washington,  Pasadena, CA 91101--1292, USA}
\footnotetext{Department of Astronomy, University of Virginia, 
Charlottesville, VA 22904--4325, USA} \footnotetext{National Optical
Astronomy Observatories, Tucson, AZ  85726--6732, USA}
\footnotetext{Department of Physics, University of Oxford, Oxford OX1 
3PU, United Kingdom} \footnotetext{NASA-Jet Propulsion Laboratory,
Pasadena, CA 91109, USA} \footnotetext{Cerro Tololo Inter-American
Observatory, La Serena, Chile} \footnotetext{School of Earth and Space
Exploration, Arizona State  University, Tempe, AZ 85287--1404, USA}
\footnotetext{SOFIA Science Center, NASA Ames Research Center, Moffett
Field, California 94035, USA\\ \hspace*{0.4cm}{$^\star$} Based on
observations with the NASA/ESA {\it Hubble Space Telescope}, obtained at
the Space Telescope Science Institute, which is operated by AURA, Inc.,
under NASA contract NAS5-26555} \hspace*{-2cm}  In spite of the rather
large distances ($\sim 50$\,kpc) involved (Hunter et al. 1995a), these
objects can be studied even in the very compact core of NGC\,2070,
referred to as R\,136, which was first resolved unambiguously by the
Hubble Space Telescope (HST) using the Faint Object Camera (Weigelt et
al. 1991; De Marchi et al. 1993). But we still know very little about
the other part of this population consisting of the fainter, low mass
main sequence (MS) and pre-MS (PMS) stars and brown dwarfs. These might
be present in even larger numbers if the tapered power law IMF proposed
recently by De Marchi, Paresce \& Portegies Zwaart (2010) for young and
very old clusters continues to the hydrogen burning limit and beyond.
Their number and spatial distribution are critical for understanding the
low mass star formation process in a cluster of very massive stars and
their role in determining the ultimate fate of the cluster itself.

The current observational situation regarding low mass stars in
NGC\,2070 is contradictory. Although there is good evidence for a
significant population of these stars, the precise nature of their
spatial and mass distributions is not well understood. Sirianni et al
(2000), using $V$- and $I$-band HST observations claim that the mass
function (MF)  flattens below $\sim 2$\,\Msolar, while Zinnecker et al.
(2002) and more recently Andersen et al. (2010), using HST observations
in the H band, find no such feature in their data down to their
completeness limit around 1\,\Msolar. So far, little can be said of what
happens below this limit, mainly because these stars in NGC\,2070 and
its compact core R\,136 are very difficult to detect and characterize
accurately. First, because of their inherent faintness especially with
respect to the high background due to the numerous much brighter stars
in the field. Second, because they scatter widely across the
colour--magnitude diagram (CMD) making accurate subtraction of field
stars and older cluster PMS stars and subdwarfs near the MS unreliable
without useful separation criteria. Finally, the patchy extinction
across the face of the nebula (e.g. Indebetouw et al. 2009) needs to be
properly accounted for in order to accurately determine their true
brightness. 

For these reasons, use of the Wide Field Camera 3 (WFC3), the new
panchromatic camera on HST, is, in principle, ideal for the task of
establishing the main characteristics of the low mass population in
NGC\,2070. With its high spatial resolution over a wide field
encompassing most of the central part of the cluster and its
unprecedented sensitivity in the near UV and IR, the WFC3 allows us to
tackle for the first time most of the problems summarised above with
some hope of resolving them. In this paper, we present the first results
of a deep survey of NGC\,2070 obtained in 2009 October as part of the
Early Release Science programme with the WFC3 camera. This project has
as its ultimate objective that of determining as accurately as possible
the properties of the low-mass initial MF in this cluster, which could
be a prototype of the much larger and distant unresolved super star
clusters in starburst galaxies. In order for this type of study to
become possible, we must first carefully identify and distinguish from
one another stars of different ages, since several generations of
objects are known to populate the 30\,Dor region (e.g. Walborn \& Blades
1997).

So far, studies of the star formation history of 30\,Dor have been
limited to the most massive stars, typically above 20\,\Msolar, and
hence to the past 10\,Myr or so, because these works rely on
spectroscopy of individual objects that cannot easily be obtained for
less massive stars (e.g. Selman et al. 1999). Clearly, access to a wider
mass range is crucial, so the goal of this paper is to understand how
star formation has proceeded for much smaller objects, i.e. stars with
masses as low as $\sim 0.5$\,\Msolar. Therefore, in this work we
concentrate on a systematic search for low-mass stars in the PMS phase,
which we can identify in a reliable way and which allow us to study how
star formation has proceeded in this area over the past $\sim 30$\,Myr.
Only then will it be possible to address the properties of the MF, which
will be the subject of future work.

The structure of this paper is as follows. We describe the observations
in Section\,2, the reduction of the data in Section\,3 and the 
photometric analysis in Section\,4. The correction for extinction and 
the resultant corrected CMDs are presented in Section\,5. The 
Hertzsprung--Russell (HR)  diagram and its analysis with recent PMS 
models are the topic of Section\,6, where we also discuss the
preliminary physical implications  of our results on the star formation
history in these regions. A summary is provided in Section\,7.

\section{Observations}

The observations discussed in this paper were obtained in 2009 October
(20 -- 27) with the WFC3 on board the HST. We refer the reader to
Dressel et al. (2010) for information about the WFC3 instrument and its
performances. Six broad-band and one narrow-band filters were used in
these observations, as shown in Table\,\ref{tab1}, where the number of
exposures and the total exposure times in each filter are also given.
Note that an extensive dithering pattern was employed, combining both
long and short exposures, in order to improve the sampling of the
telescope's point spread function (PSF), to minimise the effects of
detector blemishes, to cover the gap between the two detectors in the
field of view, and to recover the magnitudes of the brightest objects
that would otherwise saturate in long exposures.

\begin{figure*}[t]
\centering
\resizebox{14cm}{!}{\includegraphics{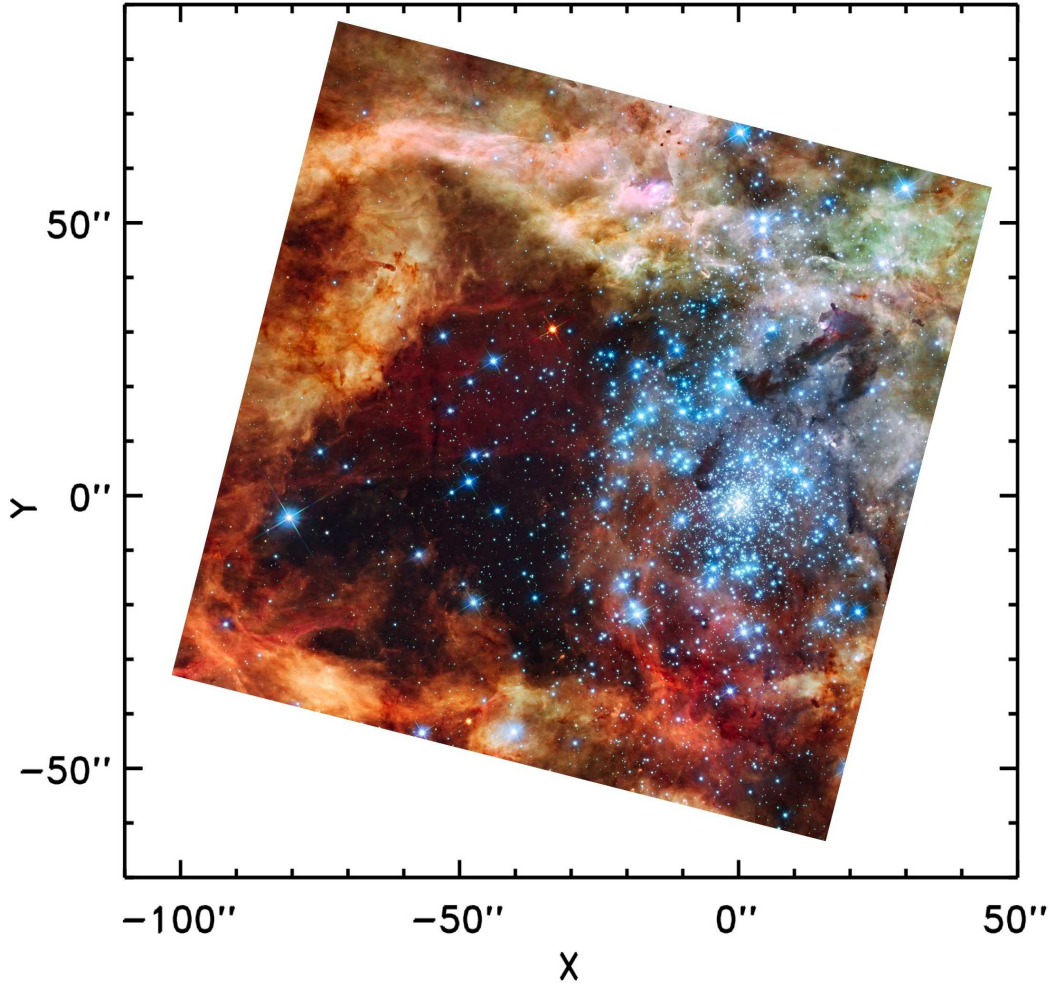}}
\caption{Colour-composite image of the central part of the observed
field.  The blue channel is obtained by averaging the F336W and F439W
filters, the F555W filter serves as green channel, the F814W filter as
red channel and the F656N filter is used as orange. { The $(0,0)$
position in this figure corresponds to the nominal centre of R\,136
(RA$=05^{\rm h}\;38^{\rm m}\;43\fs3$, DEC$=-69\arcdeg\;06\arcmin\;08\arcsec$;
J2000), with North pointing up and East to the left. At the distance of
R\,136, one parsec corresponds to $\sim 4\arcsec$.} (This image was
released by the Space Telescope Science Institute on 2009 December 15 as
part of News Release number STScI-2009-32.)} 
\label{fig0}
\end{figure*}

\begin{deluxetable}{lcr}[bh]
\tablecolumns{3}
\tablewidth{20pc}
\tablecaption{List of filter names, number of exposures and total
exposure times of the WFC3 observations of 30\,Dor.} 
\tablehead{\colhead{Filter name} & \colhead{$N_{\rm exp}$} &
\colhead{$t_{\rm tot}$}}
\startdata
F336W & 24 & $8\,659$ \\
F438W & 16 & $5\,174$ \\
F555W & 20 & $6\,892$ \\
F656N &  8 & $10\,805$ \\
F814W & 20 & $10\,700$ \\
F110W &  9 & $1\,518$ \\
F160W & 12 & $7\,816$ 
\enddata
\tablecomments{Total exposure times ($t_{\rm tot}$) are in s.}   
\label{tab1}
\end{deluxetable}

For illustration purposes, Figure\,\ref{fig0} presents a four-colour
composite image of the observed field. The colours result from assigning
different hues to each monochromatic image, as indicated in the figure
caption. The bright compact star cluster in the upper right-hand side of
the frame is R136, the core of the massive NGC\,2070 association. A
cursory glance at Figure\,\ref{fig0} already shows that the centre of
the cluster is dominated by bright blue stars, although a handful of
bright red objects are seen to the East of it, suggesting the presence
of an older generation of stars. This is consistent with the conclusions
of Selman et al. (1999), who found at least three generations of massive
stars, with the youngest and bluest members more centrally concentrated.
We will return in more detail to the spatial distribution of these
objects in Section\,4.

\section{Data reduction and photometry}

Throughout this paper we will concentrate on the observations in the
optical wavelength range, and in particular on the F555W, F656N and
F814W bands (hereafter referred to as $V$, $H\alpha$ and $I$,
respectively), whereas the entire WFC3 dataset including the
observations at IR  wavelengths will be covered in more detail in a
forthcoming paper (Beccari et al., in preparation).  

As regards the data reduction, the long and short exposures were treated
in a different way. The deep datasets were combined using the IRAF {\em
multidrizzle} task, with the appropriate calibration files and geometric
distortion coefficients for the UVIS channel of the WFC3. Given the
large number of deep frames available in the UVIS filters and  the
careful dithering  strategy adopted for these observations, we were able
to derive one single ``drizzled'' image per band, free of cosmic rays
and blemishes and with a rather sharp and well sampled PSF, ideal for
the detection of faint stars. Producing a deep, high-quality combined
image per filter is particularly useful in a field like 30\,Dor where 
stars are embedded in extensive nebulosity associated with the gas that
makes the photometry and the determination of the background more
difficult on individual images.

Star detection and identification was done by running the automated IRAF
{\em daofind} routine separately on the $V$- and $I$-band drizzled frames,
with the detection threshold set at $5\,\sigma$ above the local
background level. In order for objects to be classified as real stars,
we required that they be detected in both filters. The presence of
numerous gas filaments and irregularities in the background can cause
spurious detections, particularly in the H$\alpha$ band, where
considerable emission is present (see Figure\,\ref{fig0}. For this
reason, we visually inspected the detected stars and excluded from the
bona-fide list all the detections not clearly associated with point-like
sources. 

More specifically, we applied an unsharp-masking filter (e.g. Malin
1977), with a smoothing radius of 15 pixel meant to subtract the
low-frequency signal. Unsharp masking makes structures and filaments in
the nebular gas easier to see than in the direct image. In this way we
compared the structures and filaments with the positions of the detected
stars, in order to avoid spurious detections (see Beccari et al. 2010
for  an application of this method to the case of the massive Galactic
cluster NGC\,3603 observed with the same camera).

Once the master list was defined in this way, we proceeded to measuring
the magnitudes of each star with the IRAF {\em phot} aperture
photometry task. Since the central regions are rather crowded, we opted
for a photometric aperture of 3 pixel radius and a background annulus
enclosing the area between 4 and 7 pixel radius around the stars.
Following De Marchi et al. (1993), we took as background value the mode
of the counts within the annulus, in such a way to minimise the
contamination due to neighbouring objects. This procedure was not
applied directly on the combined drizzled frames, but rather on a subset
of those. More precisely, having at our disposal 16 deep exposures both
in the $V$ and $I$ bands, we used multidrizzle to build for each filter four
drizzled images, each comprising four individual exposures. We then ran
the IRAF phot task on each of them, using as input coordinates the
master list derived before. In this way, we could derive four
independent measurements of the magnitude for each object in both the
$V$ and $I$ band and the standard deviation of the four measurements
immediately provides the photometric uncertainty. We followed the same
procedure in all other visible bands, including H$\alpha$ for which the
four combined images comprise each two individual exposures.

The approach followed for the short exposures (ranging from a $0.5$\,s
to 30\,s in duration) is different, owing to the limited number of
exposures available in each filter. In that case we performed the
photometry on each individual flat-fielded image using PSF fitting,
having first built a proper PSF model on each frame using isolated and
well sampled stars, which was then fitted to each object in the images
by means of the standard DAOPHOT/ALLSTAR routine. The reason why we did 
not use PSF-fitting photometry also for the deep exposures is that image
combination with multidrizzle does not necessarily preserve the
uniformity of the PSF.

The two photometric catalogues obtained in this way for the long and
short exposures were independently calibrated following the specific
recipes developed by Kalirai et al. (2009) for the UVIS channel of the
WFC3. Briefly, all magnitudes were first rescaled to an aperture of 
$0\farcs4$ and then transformed into the VEGAMAG system using the
synthetic zero points given by Kalirai et al. (2009). The accuracy of
the photometry in both catalogues is very good. The typical photometric
uncertainty over the entire sample ranges from  $< 0.02$\,mag at  $V <
20$ or $I < 19$ to $\sim 0.2$\,mag at $V = 26.5$ or $I = 25.5$ (note
that in the following sections we will limit our study to objects with
the smallest uncertainties, as indicated). 

Finally, the two photometric catalogues were combined and merged so as
to avoid duplications. We used stars in the magnitude range $16 < V<
19$, where errors are small, to verify the integrity of the photometry 
by comparing the magnitudes derived for the same star in the long and 
short exposures. We found the match to be excellent, with any
differences in magnitude being comparable to the photometric
uncertainties in the short exposures ($< 0.03$\,mag). 

\begin{figure}[t]
\centering
\resizebox{\hsize}{!}{\includegraphics[bb=50 0 484 420]{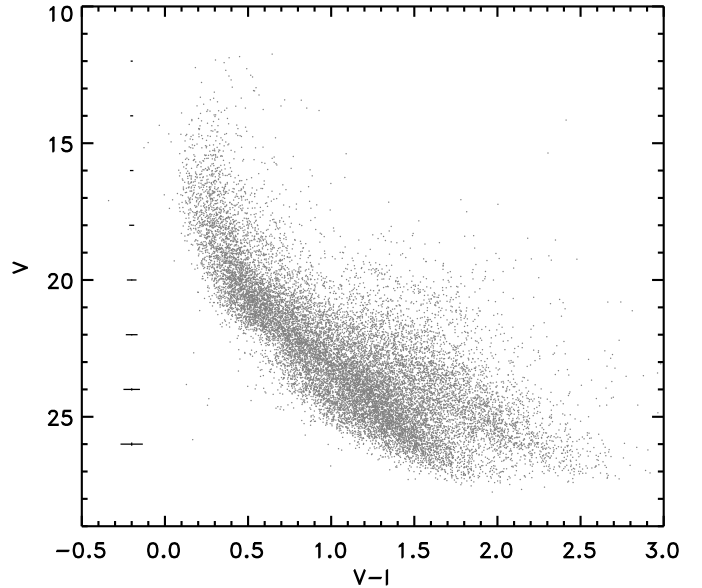}}
\caption{Colour--magnitude diagram of the 30\,Dor region covered by
these observations in the $V$ and $I$ bands. Shown are all stars with a
combined photometric uncertainty in $V$ and $I$ of less than $0.07$ mag. The
horizontal bars on the left-hand side of the figure indicate the typical
uncertainty on the $V-I$ colour as a function of magnitude.} 
\label{fig1}
\end{figure}

The final combined catalogue contains 22\,291 stars with well defined
fluxes in the $V$ and $I$ bands. 
Of these, 18\,142 objects are also detected in H$\alpha$. The CMD of 
all objects with combined photometric uncertainty in the $V$ and $I$ bands
not exceeding $0.07$\,mag is shown in Figure\,\ref{fig1}.

\section{Differential reddening}

The CMD shown in Figure\,\ref{fig1} is the deepest so far obtained for
this cluster. At $V-I \simeq 1.5$ it reaches about 3\,mag deeper than 
the CMDs previously obtained with the WFPC2 camera on board the {\it
HST} in these bands (Sirianni et al. 2000; Hunter et al. 1995a). Note
that this is not only due to the improved sensitivity of the WFC3
camera, but also to the considerably shorter exposure times in the I
band and the smaller field of view of the WFPC2 camera, which limited
the observations to the most crowded regions of the cluster. The
striking difference between the CMD of Figure\,\ref{fig1} and that of
Sirianni et al. (2000) is the prominent lower MS occupying the region
fainter than $V \simeq 22$ and bluer than $V-I \simeq 1.5$ and revealing
the presence in  this field of a considerably older stellar population
than the $\sim 3$\,Myr old stars associated with R\,136 (e.g. De Marchi
et al. 1993; Hunter et al. 1995a; Massey \& Hunter 1998). We will
discuss the age distribution of the stars in this field in Sections\,5
and 6.

The other interesting feature seen in Figure\,\ref{fig1} is the large
number of objects in the range $0.5 \lesssim V-I \lesssim 2$ that appear
brighter than the lower MS. Their location is consistent with that of
the PMS population studied by Sirianni et al. (2000; see also Andersen
et al. 2009), but their number is much larger than those detected
before. We discuss the properties of these stars in Section\,5, but we
first need to address an important issue that affects the CMD, namely
interstellar extinction.

The width of the upper MS seen in Figure\,\ref{fig1}, at $V \lesssim
20$, is much larger than the photometric uncertainty at those magnitudes
(indicated by the horizontal bars on the left-hand side of the figure),
revealing that there is a wide spread of extinction in this field.
Actually, there is general consensus that differential reddening is
present in these regions. From the analysis of high-resolution
narrow-band images, Hunter et al. (1995b) concluded that the nebulosity
in this area is not uniform even on scales of order $\sim 10\arcsec$.
However, in a subsequent study of the stellar populations in this region
Hunter et al. (1996) opted for a uniform reddening correction across the
whole field, since they concluded that uncertainties (e.g. red leak) in
the response of the WFPC2 filters did not justify a star-by-star
correction. Also Sirianni et al. (2000) used the same $A_V$ value for
all stars, although this may have lead to larger uncertainties in their
mass determination of individual PMS stars, as Selman et al. (1999) and
more recently Andersen et al. (2009) have pointed out. In order to
overcome this problem, in Section\,4.1 we will use a robust method based
on the determination of the individual reddening corrections for a large
group of bright stars (Romaniello 1998), which are then used to derive a
correction also for fainter neighbouring objects. 

The young population in this field, responsible for the upper MS in
Figure\,\ref{fig1}, has an age of $\sim 3$\,Myr (e.g. De Marchi et al.
1993; Hunter et al. 1995a) or perhaps younger ($\sim 2$\,Myr) if Wolf
Rayet stars are still burning Hydrogen (Massey \& Hunter 1998; de Koter,
Heap \& Hubeny 1998; Walborn et al. 1999). One can therefore determine
the reddening towards each of those stars by measuring their colour and
magnitude displacements with respect to a theoretical isochrone for the
appropriate age, metallicity and distance. As for the metallicity,
throughout this paper we will assume $Z=0.007$ or about one third
$Z_\odot$, since this is a typical value for the LMC (e.g. Hill,
Andrievsky \& Spite 1995; Geha et al. 1998). As regards the distance to
30\,Dor, we will use $51.4 \pm 1.2$\,kpc since this is the geometrical
distance to the neighbouring SN\,1987 object (Panagia et al. 1991, later
updated in Panagia 1999).The resulting distance modulus of $18.6$
is fully consistent with that adopted by (Walborn \& Blades 1997).

\begin{figure}[t]
\centering
\resizebox{\hsize}{!}{\includegraphics[bb=50 0 484 420]{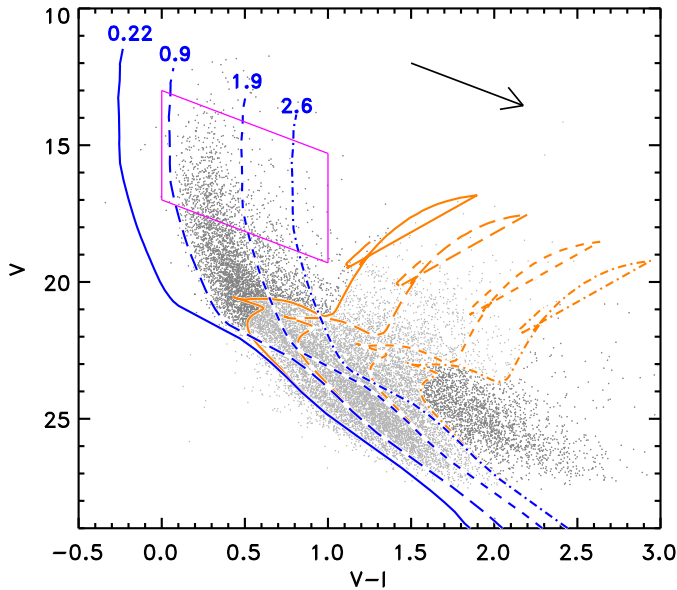}}
\caption{The CMD of Figure\,\ref{fig1} is compared here with theoretical
isochrones from Marigo et al. (2008). The lines on the left-hand side of
the figure (blue in the online version) are for an age of 3\,Myr, a
metallicity of $Z=0.007$  and a distance modulus of $18.6$. Those on the
right-hand side of the figure (orange in the online version) are for the
same distance and reddening, but for an age of 2\,Gyr. The four line
types correspond to different values of the extinction $A_V$, as
indicated. Stars whose age cannot be determined, even in an approximate
way, from their position in the CMD are shown as light dots. The box
indicates the stars in the upper MS that are used for reddening
determination. Their slanted sides are parallel to the reddening vector
indicated by the arrow (shown here for $A_V=1.55$).} 
\label{fig2}
\end{figure}

We show in Figure\,\ref{fig2} as a solid line (in blue in the online
version) an isochrone for an age of 3\,Myr and a metallicity of
$Z=0.007$, from Marigo et al. (2008), and the adopted distance modulus
of  $18.6$. Having assumed a younger age, i.e. 1 or 2\,Myr, would have
only affected the shape of the isochrone in the top $\sim 1$\,mag range,
which we will not consider in this analysis. The isochrone is also
already reddened by the amount corresponding to the intervening
absorption along the line of sight due to the Milky Way, which
Fitzpatrick \& Savage (1984) quantified in $A_V=0.22$, in turn
corresponding to $E(V-I)=0.1$.   

The other curves on the left-hand side of the figure (in blue in the
online version) represent the same isochrone reddened, for purpose of
illustration, by an additional amount of extinction, with the
long-dashed, short-dashed and dot-dashed lines corresponding
respectively to $A_V=0.9$, $1.9$ and $2.6$. As for the  extinction law,
for the portion in addition to the $A_V=0.22$ due to the Milky Way we
have used the law determined spectroscopically by Scuderi et al. (1996)
for Star\,2 in the field of SN\,1987A, which we take as representative
of the 30\,Dor region as well. The arrow in the figure shows the
corresponding reddening vector for $A_V=1.55$. The long-dashed line,
corresponding to $A_V=0.9$, forms a natural left envelope to the upper
MS at $V < 20$ and leaves very few objects in the range $0.22 < A_V <
0.9$ at those magnitudes. This is expected because of the considerable
amount of foreground extinction known to be present in the LMC towards
30\,Dor: although the $\Delta A_V=0.7$ value that we find is somewhat
larger than the $\Delta A_V \simeq 0.5$ reported by Fitzpatrick (1985),
the latter value has been determined for the brightest stars at the very
centre of 30\,Dor that generally lie in evacuated cavities of HII gas of
their own making and should therefore have a lower extinction. Note
that, at magnitudes fainter than $V \simeq 22$, where the photometric
uncertainty is still rather small (see bars in Figure\,\ref{fig1}),
there is a large number of objects between the $A_V=0.9$ and $A_V=0.22$
solid curves. In fact, the latter offers a much better fit to the lower
MS, suggesting that these stars belong to an older foreground
population. At least some of  them could be members of the much older 
LMC field, which is known to have ages in excess of $\sim 1$\,Gyr (e.g. 
Panagia et al. 2000; Harris \& Zaritsky 2004).

Interestingly, there are also a few objects that appear bluer than the
$A_V=0.22$ curve. Although they are in general consistent with our
photometric uncertainties, some of them could actually be very young
stars with a circumstellar disc seen at high inclination ($> 85^\circ$).
Objects of this type would appear somewhat bluer than their photospheric
colour due to light scattering on the circumstellar disc, as well as
several magnitudes fainter than their photospheric brightness due to
extinction along the line of sight caused by an almost edge-on disc.
However, according to the models of Robitaille et al. (2006) for the
spectral energy distribution of young stars seen at various viewing
angles, objects of this type can only account for a few percent of the
total young population, so the vast majority of stars near the
$A_V=0.22$ solid curve must belong to the field in the LMC foreground.  

The presence of a foreground population of older objects becomes even
more obvious when one looks at the other set of curves on the 
right-hand side of the figure (shown in orange in the online version),
corresponding to isochrones for an age of 2\,Gyr and the same
metallicity, distance, and reddening values used for the 3\,Myr
isochrones, also from the models of Marigo et al. (2008). The 2\,Gyr
isochrones are shown up to a mass of $\sim 1.5$\,\Msolar, corresponding
to the approximate location of the red giant clump (RC), a
characteristic  feature of the CMD of low-mass, metal-rich stars
experiencing their He-burning phase. It represents the counterpart at
high metallicity of the horizontal branch seen in CMDs of metal- poor
globular clusters. The RC usually appears as a very tight, well defined
feature in the CMD of stellar populations that could in principle be
used as a distance indicator (e.g. Stanek \& Garnavich 1998; Paczynski
\& Stanek 1998; Cole 1998; Girardi et al. 1998; Girardi \& Salaris
2001). When there is patchy absorption, like in our case, and the amount
of extinction varies from place to place, the RC becomes in practice an
elongated ellipse running parallel to the direction of the reddening
vector. This is precisely the nature of the apparent sequence of points
running diagonally between $1 < V-I < 2$ and $19 < V < 22$. This feature
can also be used to derive the extinction law (De Marchi, Panagia \&
Romaniello 2007; Panagia, De Marchi \& Romaniello 2008). 

The implications of this finding for the properties of the absorbing
material in this and other LMC fields will be addressed in detail
elsewhere (De Marchi \& Panagia 2011, in preparation), but it is
immediately clear from Figure\,\ref{fig2} that the young and old stars
present in this field have rather different extinction ranges. While
stars in the RC span the entire range $0.22 \leq A_V \leq 2.6$, there
are in practice no objects in the upper MS with $A_V < 0.9$. Therefore,
the reddening correction that we can derive using the objects in the
upper MS will only be valid for those stars and, in a statistical sense,
for low-mass stars of similar age if they have a similar spatial
distribution, but it will not be appropriate for older objects, for
which we will have to resort to an average correction. The procedure
that we followed for reddening correction is explained in detail in the
following subsection.

\subsection{Separate extinction correction for younger and older stars} 

We have used all the upper MS objects inside the box shown in
Figure\,\ref{fig2} as reference stars for our determination of the
reddening distribution  across the field, with a total of about 800
stars selected in this way. Assuming that they are all MS stars, we have
measured their displacement from the solid isochrone along the reddening
vector, whose direction is given by the arrow and the tilted sides of
the box (as mentioned above, we adopted the extinction law of Scuderi et
al. 1996). We have explicitly excluded the brightest objects where an
age difference of 1 or 2\,Myr could have an effect, albeit small, on the
shape of the isochrone. The reddening distribution ranges from
$E(V-I)=0.55$ or $A_V=1.3 $ to $E(V-I)=0.9$ or $A_V=2.1$, corresponding
respectively to the 17\,\% and 83\,\% distribution limits, with a median
value of $E(V-I)=0.67$ or $A_V=1.55$. The difference between the 17\,\%
and 83\,\% limits indicates a $\pm 1\,\sigma$ spread of $0.17$\,mag in
$E(V-I)$, indicating that reddening varies considerably in this field.

The reddening values towards each of the selected MS stars were then
used to derive a reddening correction for the objects in their
vicinities. This procedure implicitly assumes that stars projected
within a relatively small region of the field have similar extinction
values, even though no information is available on the depth of their
distribution nor on the uniformity of the absorbing material. In order
to verify the validity of this assumption, we first applied the
reddening correction to the upper MS stars themselves. Each star was
corrected using the average of the extinction values of the nearest
upper MS stars, excluding the star itself (no distance weighing was
used, since we have no information on the depth of the distribution).
After experimenting with a different number of neighbours (namely 1, 5,
10, and 20), we concluded that using the 5 closest neighbours produces
the tightest upper MS. This choice corresponds to using reference stars
within a radius of less than $10\arcsec$, with a typical effective
radius of $\sim 7\farcs5$. After reddening correction, the $\pm
1\,\sigma$ spread in the upper MS drops to $0.10$\,mag and the upper MS
itself appears considerably tighter in the CMD than in the raw data (see
Figure\,\ref{fig6}). This result convinced us that the method, albeit
statistical in nature, can be applied to other objects in the field,
provided that they have an age and spatial distribution consistent with
that of the upper MS stars.

The CMD and isochrones in Figure\,\ref{fig2} help us to identify stars
of similar age. This is the case of MS stars roughly brighter than $V=
20$  and bluer than $V-I = 1$ and of young PMS objects roughly redder
and fainter than the dot-dashed 2\,Gyr isochrone (these objects are
shown as dark dots in the figure). However, the CMD alone does not allow
us to assign an age to objects falling in the region between the solid
and dot-dashed 2\,Gyr isochrones (shown as light dots), since they could
be PMS stars of various ages, low-mass MS stars, or red giants.
Therefore, in order to identify bona-fide young objects in this region
we have to resort to a more reliable youth indicator, namely a strong
H$\alpha$ excess emission. This feature is characteristic of low-mass
star formation and is attributed to the accretion process, whereby
gravitational energy, released by infalling matter, ionises and excites
the surrounding gas  (e.g. K\"onigl 1991; Shu et al. 1994). 

A detailed account on how to reliably detect these objects using
multi-colour photometry is given in De Marchi et al. (2010) and De
Marchi et al. (2011). The method, already employed also by Beccari et
al. (2010) to search for objects with H$\alpha$ excess in NGC\,3603,
will be used in Section\,5 for an accurate selection of bona-fide PMS
stars after reddening correction. Briefly, the method uses the median $V
- H\alpha$ dereddened colour of stars with small ($< 0.05$\,mag)
photometric uncertainties in the three bands ($V, I$ and $H\alpha$) to
define a reference template that is used to identify objects with excess
H$\alpha$ emission. However, as De Marchi et al. (2010) have shown, even
before reddening correction  the colour--colour diagram $V-H\alpha$,
$V-I$ provides a robust identification of stars with H$\alpha$ excess,
since in these bands the reddening vector runs almost parallel to the
median photospheric colours of normal stars. Therefore, following that
procedure, we looked initially for stars whose $V-H\alpha$ colour
exceeds by at least three times the photometric uncertainty the
$V-H\alpha$ colour of normal stars (i.e. those with no excess) with the
same $V-I$ colour. A more stringent selection will be applied after
reddening correction (see Section\,5). 

\begin{figure}[t]
\centering
\resizebox{\hsize}{!}{\includegraphics[bb=50 0 484 420]{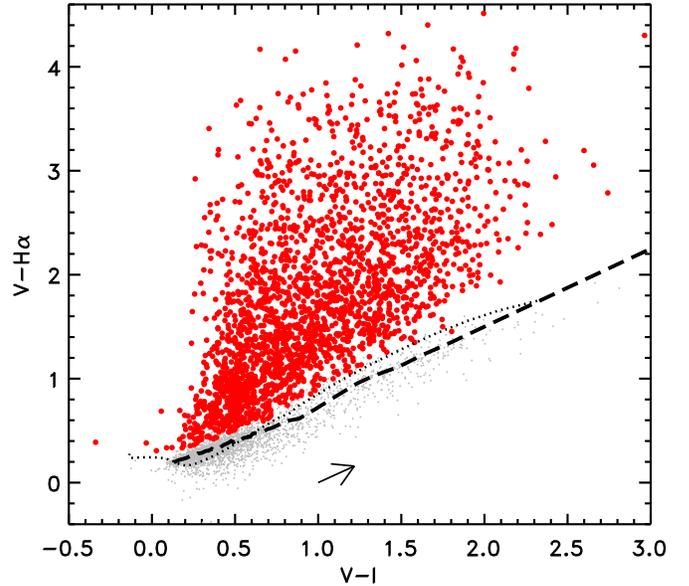}}
\caption{Colour--colour diagram used to identify stars with H$\alpha$
excess emission. Small dots represent all stars with combined
photometric error in $V$, $I$ and $H\alpha$ of less than $0.05$\,mag.  The
dashed line represents the running median $V-H\alpha$ colour,  obtained
with a box-car size of 100 points, whereas the thin dotted line  shows
the model atmospheres of Bessell et al. (1998) reddened by
$A_V=0.5$\,mag according to the extinction law of Scuderi et al.
(1996).  The arrow in the figure corresponds to the reddening vector for
$A_V=0.5$ for the specific bands used in this study. A total of 2\,374
objects with a $V-H\alpha$ excess larger than $3\,\sigma$ are indicated
with thick dots (in red in the online version).  } 
\label{fig3}
\end{figure} 

The PMS candidates selected in this way, a total of 2\,374, are shown as
thick dots in Figure\,\ref{fig3}, where the dashed line represents the
median photospheric $V-H\alpha$ colour as a function of $V-I$. For
comparison, we also show as a thin dotted line the corresponding colours
derived from the model atmospheres of Bessell, Castelli \& Pletz (1998)
for stars with effective temperatures in the range 3\,500 \,K $\le
T_{\rm eff} \le $ 40\,000\,K, surface gravity $\log g=4.5$ and a
metallicity index $[M/H]=-0.5$, as appropriate for the LMC  (Dufour
1984). The theoretical colours were calculated for the specific filters
of the WFC3 camera and were reddened by $A_V=0.5$, following the
extinction law of Scuderi et al. (1996) mentioned above for the purpose
of comparison with the data. Note that, due to the still preliminary
calibration of the WFC3 photometric system in the H$\alpha$ band, we had
to apply a small correction to the $V-H\alpha$  colour of $0.05$ mag to
force the $V-H\alpha$ colour of Vega to be 0 at $V-I=0$, as is
appropriate for the VEGAMAG photometric system.

\begin{figure}[t]
\centering
\resizebox{\hsize}{!}{\includegraphics[bb=20 0 484 420]{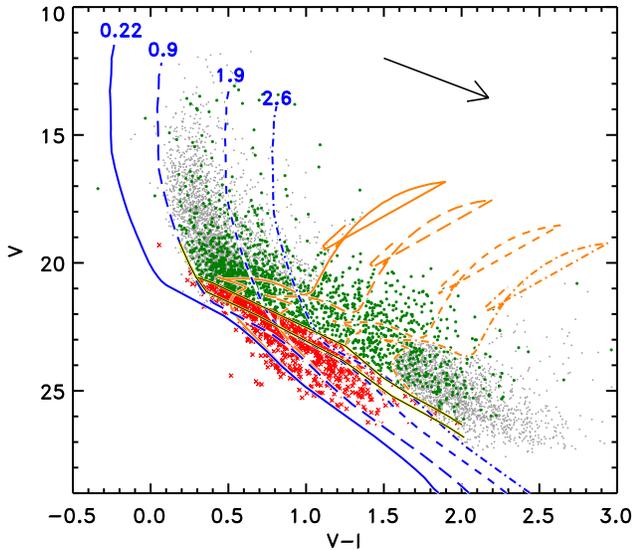}}
\caption{Same as Figure\,\ref{fig2}, but in addition we show the 
isochrones for PMS ages of 5 and 10\,Myr (thin solid lines) for $A_V=0.9$.
Stars with H$\alpha$ excess emission located above and below the 5\,Myr
isochrone are indicated respectively with thick dots (green in the
online version) and crosses (red in the online version) and are
approximately younger and older than 5\,Myr. The arrow corresponds to
the reddening  vector for $A_V=1.55$, which is the median reddening
value of upper MS stars.} 
\label{fig4}
\end{figure}

The objects with H$\alpha$ excess emission are also shown as thick
symbols in the CMD of Figure\,\ref{fig4}. Besides the 3\,Myr and 2\,Gyr
isochrones, we show there also the isochrones for PMS stars of age
5\,Myr (upper thin line) and 10\,Myr (lower thin line) from the models
of Degl'Innocenti et al. (2008) for the assumed LMC metallicity
($Z=0.007$) and having taken $A_V=0.9$ for the
reddening.\footnote[1]{Note that the ages of 5 and 10\,Myr of these PMS 
isochrone are not counted from the ZAMS but from the birth line (see
Palla \& Stahler 1993 for a definition). For instance, the 5\,Myr PMS
isochrone  meets the MS isochrone for age 3\,Myr (counted from the ZAMS)
at $V\simeq 21$, corresponding to $\sim 2.5$\,\Msolar. At this stage of
PMS evolution,  any more massive stars have already reached the  MS.}
These isochrones allow us to distinguish in an approximate way between
younger and older PMS stars (a more accurate age determination requires
the reddening correction and will be possible only at the end of this
process) and to study their spatial distribution. Using the 5\,Myr PMS
isochrone as a reference, we have marked as thick dots and crosses
respectively the objects above and below that line, and hence
approximately younger and older than 5\,Myr.

\begin{figure}[t]
\centering
\resizebox{\hsize}{!}{\includegraphics[bb=20 0 484 420]{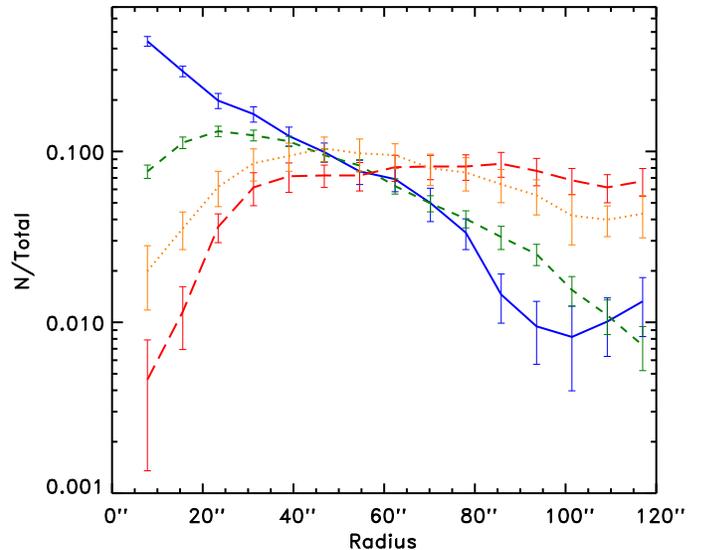}}
\caption{Radial distribution of upper MS stars (solid line), of 
low-mass stars with H$\alpha$ excess with ages younger than $\sim
5$\,Myr (short-dashed  line) and older than $\sim 10$\,Myr (long-dashed
line), as per Figure\,\ref{fig4}. The dotted line corresponds to stars
with ages intermediate between 5 and 10\,Myr. The curves are normalised
by the number of objects in each sample (respectively 812, 1637, 433 and
301), but the solid line is further rescaled by a factor of $0.6$ for an
easier comparison with the others. Error bars correspond to the Poisson
uncertainty on the number counts.} 
\label{fig5}
\end{figure}

In Figure\,\ref{fig5} we show the radial distribution of these low-mass
stars with H$\alpha$ excess, drawn from the nominal centre of R\,136.
The short-dashed line corresponds to stars younger than $\sim 5$\,Myr,
the long-dashed line to stars older than 10\,Myr, while the distribution
of  upper MS stars is shown by the solid line, rescaled  by a factor of
$0.6$ to make the comparison easier. Younger and older PMS stars have
markedly different distributions: older stars have a rather uniform
distribution and are almost absent inside a radius of $\sim 30\arcsec$,
while younger objects show a density gradient that follows remarkably
well that of upper MS stars down to $\sim 10\arcsec$, where photometric
incompleteness starts to dominate as indicated by the rapidly increasing
photometric uncertainties in these regions. The choice of which
approximate age (5\,Myr or 10\,Myr) to adopt for separating ``younger''
and ``older''  stars is somewhat arbitrary, but the dotted line in
Figure\,\ref{fig5},  corresponding to stars with age in the range $5 -
10$\,Myr, suggests  that objects with H$\alpha$ excess older than $\sim
5$\,Myr have a  radial distribution remarkably different from that of
upper MS stars. This  is perhaps not surprising, considering that 5\,Myr
is about twice the age of the massive members of R\,136 ($2-3$\,Myr; De
Marchi et al. 1993; Hunter et al. 1995a; Massey \& Hunter 1998; de Koter
et al. 1998; Walborn et al. 1999).   

We will discuss these differences in the radial distribution in more
detail in Section\,6. For now, Figure\,\ref{fig5} confirms that it is
appropriate to use the reddening values derived from upper MS stars also
for similarly young low-mass objects, but it suggests that this type of
correction would not be suitable for older stars outside of the inner
$\sim 50\arcsec$. As Figure\,\ref{fig5} implies, these objects are
uniformly distributed over the field and are thus consistent with being
in the foreground of R\,136. In order to obtain at least an estimate of
their reddening, we can use the stars with H$\alpha$ excess that appear
bluer than the isochrone for $A_V=0.9$, i.e. the long dashed line in
Figures\,\ref{fig2} and \ref{fig4}. 

As mentioned earlier, $A_V=0.9$ is the lower limit to the reddening
towards the upper MS stars of R\,136, so objects bluer than the
long-dashed line in those figures are most likely in the foreground. A
total of 144 stars satisfy this condition and, assuming that they are in
the late stages of PMS evolution, i.e. close to  the MS, we derive for
them a median reddening value of $A_V=0.5$. If these objects were in a
considerably earlier stage of their PMS evolution, this procedure would
give us an upper limit to their reddening. On the other hand, the fact
that they are uniformly distributed over the field indicates that they
cannot be as young as the other much more concentrated objects with
H$\alpha$ excess (see Figure\,\ref{fig5}). We therefore take $A_V=0.5$
as a representative value for the reddening towards all these stars and
in general for all objects with H$\alpha$ excess shown as crosses in
Figure\,\ref{fig4}. 

We can now correct for extinction the magnitudes of all objects shown in
Figure\,\ref{fig4} in the following way. All stars located within
$50\arcsec$ of the centre of R\,136 are corrected individually using the
average extinction value of the nearest five upper MS stars. As regards
objects outside $50\arcsec$, those with an age lower than $\sim 5$\,Myr
(i.e. thick and thin dots in Figure\,\ref{fig4}) are corrected in the
same way, whereas older stars with excess (crosses in
Figure\,\ref{fig4}, red in the online version) are corrected using the
average $A_V=0.5$ value. No correction is attempted for other objects,
since we cannot determine  their ages on the basis of the CMD alone, and
they are not considered further in this work. 

Note that when the magnitude of an object is corrected for extinction
using its nearest neighbours, it is assumed that the placement of the
object with respect to the absorbing material is similar to that of the
reference stars in its vicinities. This process is accompanied by some 
statistical fluctuations, since as we have seen there is an intrinsic 
spread of $0.10$\,mag in the $E(V-I)$ distribution of our upper MS 
reference stars. However, the uncertainties that this assumption can
introduce in the final magnitudes of the stars are relatively small. The
typical standard deviation of the $E(V-I)$ values of the five nearest
neighbours is $0.12$\,mag and, more generally, over the entire field
these statistical uncertainties are not expected to exceed the reddening
spread measured for the upper MS reference stars before reddening
correction, i.e. $0.17$\,mag in $E(V-I)$ or  $\sim 0.35$\,mag in $A_V$.

The CMD corrected for extinction in this way is shown in
Figure\,\ref{fig6}. It must be understood that it does not cover all
objects that were observed, but only those for which we could determine
a reliable correction for reddening. The symbol types are the same as
used in Figure\,\ref{fig4}, the solid line represents the theoretical
isochrone from Marigo et al. (2008) for a MS age of 3\,Myr, a
metallicity of $Z=0.007$ and a distance modulus of $18.6$, whereas the
dashed line corresponds to the isochrone of Degl'Innocenti et al. (2008)
for a PMS age of 5\,Myr and the same metallicity and distance modulus.
The PMS isochrone of Degl'Innocenti et al. (2008) has been translated 
from the theoretical plane ($\log L$, $\log T_{\rm eff}$) to the plane
of the observations using the model atmospheres of Bessell et al.
(1998)  folded through the specific WFC\,3 filters employed here. The
PMS isochrone meets the MS at $V_0 \simeq 19.5$ and it is reassuring
that the two curves are in excellent agreement at brighter magnitudes,
since they are based on independent models and different input
physics.   

\begin{figure}[t]
\centering
\resizebox{\hsize}{!}{\includegraphics[bb=50 0 484 420]{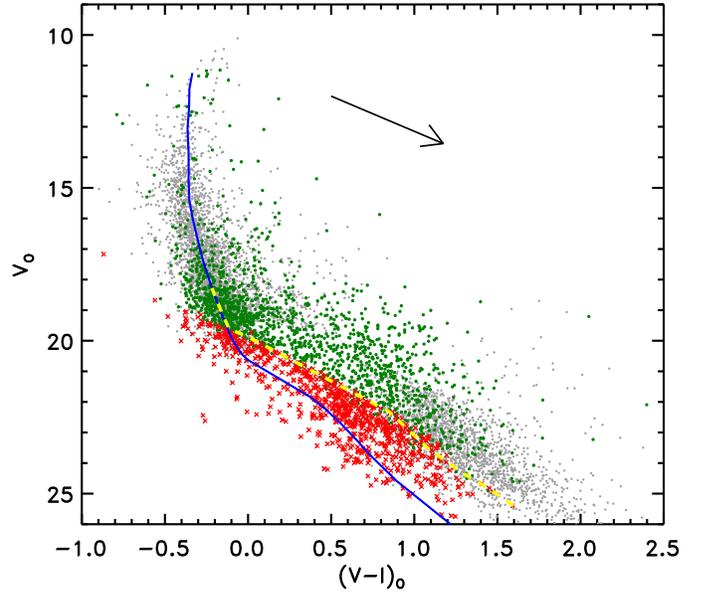}}
\caption{CMD corrected for reddening. Only stars for which a reddening
correction could be derived are shown. The symbol types are the same as 
used in Figure\,\ref{fig4} and so is the arrow, for $A_V=1.55$. The
solid line represents the theoretical  isochrones from Marigo et al.
(2008) for a MS age of 3\,Myr, a metallicity  of $Z=0.007$ and a
distance modulus of $18.6$, while the dashed line corresponds to the
isochrone of Degl'Innocenti et al. (2008) for a PMS age of 5\,Myr and
the same metallicity and distance modulus.}
\label{fig6}
\end{figure}

\section{Pre-main sequence stars}

The CMD corrected for extinction in Figure\,\ref{fig6} reveals that at
dereddened $V_0$ magnitudes fainter than $19$ there is a conspicuous
number of objects that are above (i.e. brighter than) the MS and, as
such, they are consistent with a population of PMS stars. These objects
extend down to $V_0 \simeq 26$, or more than three magnitudes deeper
than the limit reached by Sirianni et al. (2000) and Andersen et al.
(2009), who detected and studied the brightest members of this
population. 

As already mentioned in Section\,4, having corrected the photometry for
extinction allows us to better clarify and confirm the PMS nature of 
these stars by looking at the specific signatures of the mass accretion 
process that is supposed to accompany this evolutionary phase. The
accretion process is believed to be the cause of the strong excess
emission observed in these objects (e.g. K\"onigl 1991; Shu et  al.
1994). For example, the presence of a strong H$\alpha$ emission line
(with equivalent width in excess of $\sim 10$\,\AA) in young stellar 
objects is normally interpreted as the signature of ongoing accretion
(e.g. Feigelson \& Montmerle 1999; White \& Basri 2003). Note that,
although it is customary to use negative equivalent widths for emission
lines, in this work we will use $W^E_{\rm eq}(H\alpha)$ to refer to the
equivalent width of the H$\alpha$ emission line, defined as a positive
quantity. 

In Section\,4 we used this method for a preliminary coarse
identification of stars with H$\alpha$ excess emission, but here we
employ very conservative criteria to refine the selection of bona-fide
PMS stars. In particular, we now limit the search only to stars with a
combined photometric uncertainty in the $V, I$ and $H\alpha$ bands of
less than $0.1$\,mag and fainter than $V_0=17$ (corresponding to $\sim
300$\,L$_\odot$) and we look for objects with dereddened
$(V-H\alpha)_0$ colour excess at least four times larger than the
photometric uncertainty in those bands (equivalent to a $4\,\sigma$
detection limit). 

In Section\,4 we used as a reference template the median photospheric
colours of stars with small photometric uncertainty, the majority of
which are older MS stars and have no H$\alpha$ excess. This approach
cannot be followed here since many of the objects for which we could 
compute a reddening correction have H$\alpha$ excess. As a reference
template we use instead the theoretical photospheric colours derived
from model atmospheres. As Figure\,\ref{fig3} shows, the models of
Bessell et al. (1998) for surface gravity $\log g = 4.5$ and metallicity
$[M/H] = -0.5$ are in excellent agreement with the observed colours.
This was also shown to be the case for stars in the field of SN\,1987A
(De Marchi et al. 2010) and in NGC\,346 (De Marchi et al. 2011).

Besides selecting stars with excess emission above the $4\,\sigma$
level, we also imposed constraints on the equivalent width of the
H$\alpha$ emission, $W^E_{\rm eq}(H\alpha)$. Following the procedure
outlined in De Marchi et al. (2010) to derive $W^E_{\rm eq}(H\alpha)$
from the observed colours, we further restricted the selection to
objects with $W^E_{\rm eq}(H\alpha) > 20$\,\AA\, in order to avoid
contamination by stars with significant chromospheric activity.
Furthermore, since at temperatures $T_{\rm eff}  \ga 10\,000$\,K or
colours $(V-I)_0 \la 0$ the sample could be contaminated by Be stars
that are evolving off the MS, for those objects we set a more stringent
condition, namely $W^E_{\rm eq}(H\alpha) > 50$\,\AA. In a survey of
about 100 stars of type Be in the Galaxy (Cot\'e \& Waters 1987), only
one star is found with $W^E_{\rm eq}(H\alpha) > 50$\,\AA\ and the
largest majority have values in the range from $4$\,\AA\ to $30$\,\AA.

A total of 1\,159 objects satisfy these conservative detection
conditions: we will take them hereafter as bona-fide PMS stars.
Obviously, less stringent limits would result in a larger number of PMS
stars, albeit with lower significance, but it must be understood
that, even without changing the selection criteria, the actual number of
PMS stars in this field must be considerably larger. This is because our
detection and classification relies on the presence of H$\alpha$ excess
emission at the time of the observations, but very young populations are
known to show large variations in their H$\alpha$ emission over hours or
days (e.g. Fernandez et al. 1995; Smith et al. 1999; Alencar et al.
2001), with only about one third of them at any given time being active
H$\alpha$ emitters with $W^E_{\rm eq}(H\alpha) > 10$\,\AA\, (Panagia et
al. 2000; De Marchi et al. 2010; De Marchi et al. 2011). 

Therefore, the 1\,159 stars identified as bona-fide PMS objects are
those that were undergoing active mass accretion and therefore showed
H$\alpha$ excess emission at the time of these observations. On the
other hand, since the duty cycles of any two stars are completely
independent of one another, the sample of bona-fide PMS stars identified
in this way is representative of the entire PMS population in this
field. Hereafter, unless otherwise indicated, when talking about PMS
stars we will refer exclusively to the 1\,159 bona-fide PMS objects
identified through their H$\alpha$ excess emission.

\begin{figure}[t] 
\centering
\resizebox{\hsize}{!}{\includegraphics[bb=40 0 484 420]{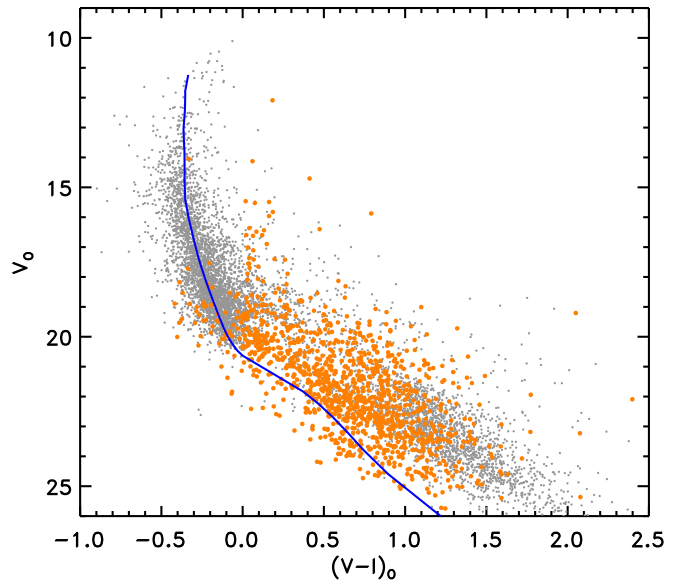}} 
\caption{Thick dots mark bona-fide PMS stars in the CMD corrected for
extinction. The majority occupy the region above the MS where young PMS
objects are expected, but there are several located at or near the MS,
suggesting an older age. The solid line is the same as in
Figure\,\ref{fig6}.}  
\label{fig7} 
\end{figure}

We show in Figure\,\ref{fig7} their locations in the CMD (thick dots).
The majority occupy the region above the MS where young PMS objects are
expected, but there are many located at or near the MS, suggesting an
older age. A similar situation has already been noticed in other star
forming regions, namely the SN\,1987A field in the LMC (De Marchi et
al.  2010), NGC\,3603 in the Milky Way (Beccari et al. 2010), and
NGC\,346 in the Small Magellanic Cloud (De Marchi et al. 2011). 

The skeptical reader might be concerned that the H$\alpha$ excess
emission of many of the objects near the MS could in reality be due to
nebular contamination of the gas present in the field. However, as
mentioned in Section\,3, we specifically excluded from our photometry
all objects for which the photometric aperture would be contaminated by
gas filaments. If the filament completely fills both the core aperture
used to measure the flux of the star and the surrounding background
annulus, it will not alter the photometry of the star, since it will be
equivalent to a uniform background. Obviously, if the background annulus
does not properly sample the contamination, for instance because the
filament only covers part of it, the emission present in the gas may be
erroneously associated to the object (for an illustration of this effect
see Beccari et al. 2010). Precisely for this reason, as discussed in
Section\,3, we have removed from the initial star lists all objects in
the vicinity of a filament. Therefore, we are confident that the
H$\alpha$ excess emission of our bona-fide PMS stars is an intrinsic
feature of those  objects and we must interpret their distribution
across the CMD as a sign of an age spread.

\begin{figure}[t]
\centering
\resizebox{\hsize}{!}{\includegraphics[bb=40 0 484 420]{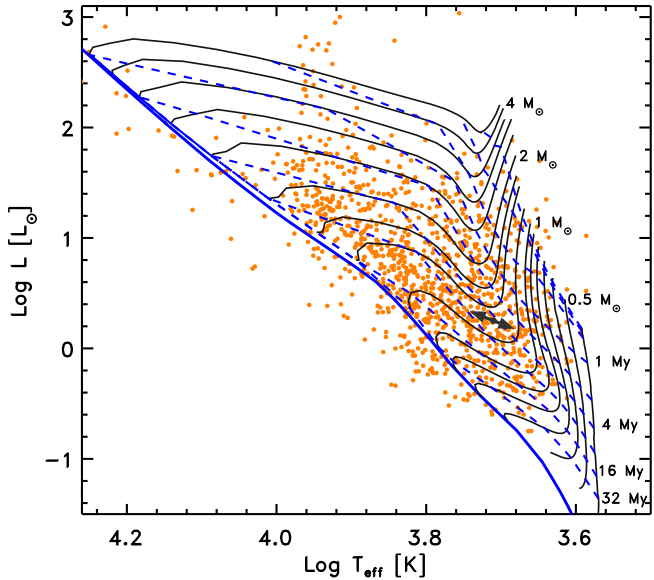}}
\caption{HR diagram of the bona-fide PMS stars. The
solid lines correspond to the PMS evolutionary tracks of Degl'Innocenti
et al. (2008) for $Z=0.007$ and masses of $0.4$, $0.5$, $0.6$, $0.7$, 
$0.8$, $0.9$, $1.0$, $1.2$, $1.5$, $1.7$, $2.0$, $2.5$, $3.0$, $3.5$, 
$4.0$ and $4.5$\,\Msolar, from bottom to top. The dashed lines indicate
the PMS isochrones from the same models for ages or $0.125$, $0.25$,
$0.5$, 1, 2, 4, 8, 16 and 32\,Myr, from right to left. The leftmost
solid line represents the ZAMS of Marigo et al. (2008) for $Z=0.007$
and, for $L>L_\odot$, it agrees rather well with the 32\,Myr PMS
isochrone. The good match between the two independent sets of models
is also confirmed by the fact that the PMS evolutionary tracks end on
the ZAMS. The double arrow at $\log T_{\rm eff} \simeq 3.7$ and $\log L
\simeq 0.2$ indicates the effect of the $\pm 1\,\sigma$ reddening 
uncertainty of $\Delta E(V-I)= \pm 0.12$\,mag.} 
\label{fig8}
\end{figure}

\section{Star formation over time}

More quantitative information on the age of these objects can be
obtained from the comparison with PMS isochrones, shown in
Figure\,\ref{fig8}, where the effective temperatures ($T_{\rm eff}$) and
luminosities ($L$) of these stars are arranged in the
HR diagram. The value of $T_{\rm eff}$ was
derived for each star directly from the dereddened $V-I$ colour, since
this is an excellent index for temperature determinations in the range 
$4\,000 - 10\,000$\,K (e.g. von Braun et al. 1998; Bessell et al. 1998).
For the conversion we made use of the models of Bessell et al. (1998)
already discussed in Section\,4. As for the bolometric luminosity, it
was obtained from the dereddened $V$ magnitude and the same models,
having assumed a distance modulus of $18.6$.

The solid lines in the figure corresponds to the PMS evolutionary track
from the Pisa group (Degl'Innocenti et al. 2008; Tognelli et al. 2011)
for metallicity $Z=0.007$. The tracks correspond to stellar masses from
$0.4$\,\Msolar to $4.5$\,\Msolar and reveal the presence of bona-fide
PMS stars over the entire mass range. The isochrones (dashed lines)
correspond, from right to left, to ages ranging from $0.125$\,Myr
through to 32\,Myr with a constant age step of a factor of 2, selected
in such a way that our photometric uncertainties are slightly smaller
than the typical separation between the isochrones. PMS objects are seen
across the entire age range.

\subsection{Age distribution of PMS stars}

The comparison of the HR diagram with PMS evolutionary tracks and 
isochrones allows us to quantify the age spread of the PMS stars that
was already evident in the CMD. Individual stellar masses and ages were
derived via interpolation over a finer grid than the one shown in the
figure. We followed the interpolation procedure developed by Romaniello
(1998), which does not make assumptions on the properties of the
population, such as the functional form of the IMF. On the basis of the
measurement errors, this procedure provides the probability distribution
for each individual star to have a given value of the mass and age (the
method is conceptually identical to the one presented recently by Da Rio
et al. 2010). 

An important aspect to consider is the uncertainty associated with the
age determination of our PMS stars from their location in the HR
diagram. Besides photometric errors in the $V$ and $I$-band data (which
are always less than $0.07$\,mag), the major sources of uncertainty are
the effects of reddening and the comparison with stellar models. As
regards the latter, De Marchi et al. (2010) show that, excluding
possible inaccuracies in the models input physics and interpolation
errors, the largest source of uncertainty comes from the use of models
that do not properly describe the stellar population under study, e.g.
because of the wrong metallicity, and from differences between models of
various authors. They conclude that systematic uncertainties of order
20\,\% are to be expected for the mass and possibly higher for the age. 

\begin{figure}[t] 
\centering
\resizebox{\hsize}{!}{\includegraphics[bb=20 0 484 390]{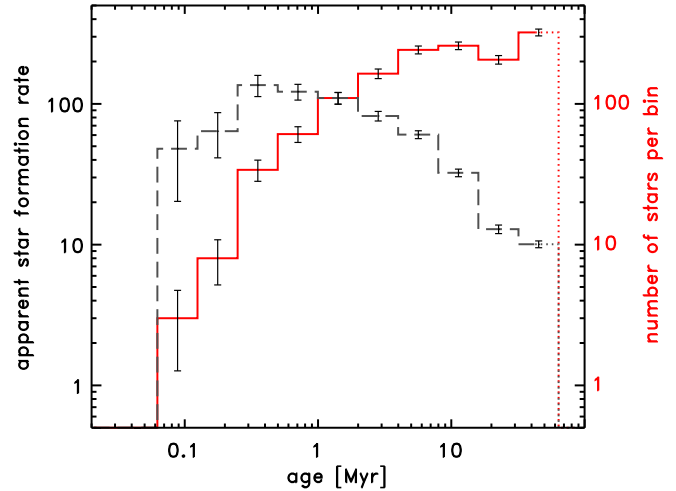}}
\caption{Histogram of the age distribution of PMS stars. Ages are binned
using a constant logarithmic step (a factor of 2). The solid line gives
the number of stars inside each age bin, whereas the dashed  line
provides an apparent value of the star formation rate in units of stars
per Myr.}  
\label{fig9} 
\end{figure}

As for reddening, dust extinction systematically displaces stars in the
HR diagram towards lower luminosities and effective temperatures.
Therefore, if the photometry of a PMS star is not corrected for
reddening, its age will typically be underestimated. A study currently
being conducted (De Marchi \& Panagia, in prep.) shows that for
reddening $E(V-I) < 1$ there is a rather linear relationship of the type
$E(V-I) \simeq \log (t_r/t_0)$, where $t_r$ and $t_0$ are the ages
before and after reddening correction. For example, as discussed in
Section\,4, the spread of the upper MS in Figure\,\ref{fig1} corresponds
to a reddening spread of $\pm 0.17$\,mag in $E(V-I)$. Therefore, if we
had used the average reddening value of $A_V=1.55$ to  correct the
magnitudes of all young stars, instead of using the values of the
nearest neighbours, we would have introduced an age uncertainty of a
factor of $\sim 1.5$. Similarly, if reddening correction for older stars
with excess (i.e. those indicated by crosses in Figure\,\ref{fig4}) had
been done using the average $A_V$ value of the nearest upper MS
neighbours instead of $A_V=0.5$, the derived ages would be typically
twice as old. As a further example, the double arrow in
Figure\,\ref{fig8} indicates the amount and direction of the
displacement in the HR diagram caused by the typical uncertainty of
$\pm 0.12$\,mag in $E(V-I)$ associated with our reddening correction
(see Section\,4.1). In the specific case shown, the corresponding
uncertainty on the age is a factor of $\sim 1.5$ while the uncertainty
on the mass is less than 15\,\%.

This analysis suggests that the typical uncertainty on the ages that we
derive is less than a factor of 2. For this reason, the age distribution
of PMS stars is shown in  Figure\,\ref{fig9} using as bin size a
constant logarithmic step of a factor of 2, which does not suffer from
the relative age uncertainties. The solid line gives the number of stars
inside each age bin as a function of time, whereas the dashed  line
provides an apparent value of the star formation rate in units of  stars
per Myr (i.e. the number of stars in each bin divided by the size of the
bin). 

\begin{figure*}[t]
\centering
\resizebox{14cm}{!}{\includegraphics[bb=0 0 430 410]{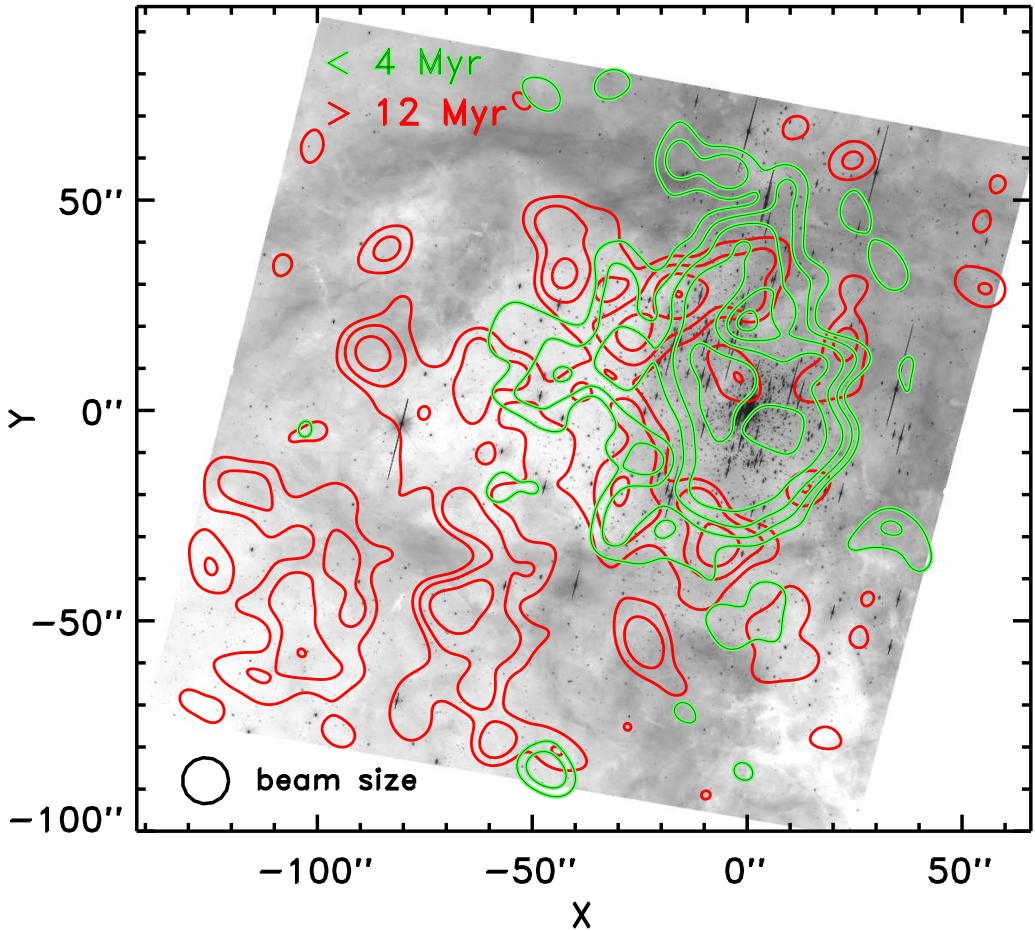}}
\caption{Contour lines showing the spatial density distribution of 
younger ($<4$\,Myr; green) and older ($>12$\,Myr; red) PMS stars,
overlaid on a negative H$\alpha$ image of the region. The $(0,0)$
position in this figure corresponds to the nominal centre of R\,136
with North pointing up and East to the left, like in
Figure\,\ref{fig0}, but here we show the entire extent of the observed
field. The contour plots have  been obtained after Gaussian smoothing
with a beam size of  $\sigma=4\arcsec$ or $\sim 1$\,pc, as indicated by
the circle in the lower left corner. The step between contour levels is
constant and corresponds to a factor of $1.5$.} 
\label{fig10}
\end{figure*}

The dashed line seems to suggest a rather uniform star formation rate
over the past 4\,Myr, of order $\sim 200$\,\Msolar\,Myr$^{-1}$ (the mass
of our PMS stars younger than 4\,Myr ranges from $1.1$\,\Msolar to
$3.2$\,\Msolar, respectively the 17 and 83 percentiles, with an average
of 2\,\Msolar), and a progressive drop at earlier times. Obviously, this
is necessarily a lower limit to the star formation rate, since our
measurements only account for PMS stars in the range  $\sim 0.5 -
4.0$\,\Msolar that had H$\alpha$ excess emission at the time of  the
observations. Furthermore, since the fraction of PMS stars in an active
state of accretion is expected to decrease with age (see De Marchi,
Panagia \& Sabbi 2011), the apparent drop in the star formation beyond
2\,Myr can at least in part be explained.

Keeping in mind that we can only provide a lower limit to the star
formation rate, near the peak of  the distribution, at $\sim 1$\,Myr,
this limit corresponds to  $\sim 330$\,\Msolar\,Myr$^{-1}$ (the average
mass of those objects is $2.3$\,\Msolar), while at $\sim 16$\,Myr it
drops by over an order of magnitude to $\sim 10$\,\Msolar\,Myr$^{-1}$
(average mass $1.3$\,\Msolar). A more detailed study of the star
formation rate of 30 Dor will be the topic of a future paper, but the
histograms in Figure\,\ref{fig9} already show quite convincingly that
star formation has been present in this region for an extended period of
time, exceeding 30\,Myr. 

\subsection{Spatial distribution of PMS stars}

From a spectroscopic study of massive stars ($> 40$\,\Msolar) within a
radius of $\sim 15\arcmin$ of R136, Selman et al. (1999) concluded that
recent star formation in this region shows at least three distinct
bursts, respectively $\sim 5, \sim 2.5$ and $\sim 1.5$\,Myr ago, with
increasing strength. They also found that the younger generation is
considerably more concentrated than the others and is responsible for
most of the star-formation within 6\,pc or $\sim 25\arcsec$ of the
cluster centre. It is therefore useful to compare this picture with the
spatial distribution of our PMS stars.

Although we do not have the age resolution needed to establish
unambiguously whether the formation has been continuous during the past
30\,Myr or whether repeated bursts have occurred, we can look at the 
current age distribution for clues as to the relationships between 
stars of different ages. The age distribution of these PMS stars is
characterised by a median of $\sim 7$\,Myr, with about $35\,\%$ of the
stars being younger than $\sim 4$\,Myr and an equal number older than
$\sim 12$\,Myr. Hereafter we will refer to these two groups, comprising
415 objects each, as ``younger'' and ``older'' PMS stars, respectively.
Note that, with a typical age uncertainty not exceeding a factor of 2,
these two groups are well separated in age, since the former has a
median age of $\sim 2$\,Myr and the latter $\sim 18$\,Myr.

Interestingly, also their projected spatial density distributions are
very different. They are shown in Figure\,\ref{fig10} by contour lines
of different colour, with logarithmic scaling, plotted over an image of
the field in the  H$\alpha$ band. The contour plots have been derived
after smoothing the distribution with a Gaussian beam with size
$\sigma=4\arcsec$ or $\sim 1$\,pc, as indicated by the circle in the
lower left corner of the figure. The lowest contour level corresponds to
a local density of PMS stars twice as high as the average PMS stars
density over the entire field. The step between levels corresponds to a
factor of $1.5$. 

There is a pronounced difference in the spatial distribution of younger
and older PMS stars. The former are concentrated near the centre of the
R\,136 cluster, where they form a density plateau, while older PMS stars
occupy preferentially the region to the east of R\,136 and are remarkably 
absent from the cluster centre. Since these stars have a rather uniform
distribution across the field, we should expect to find them also near
the location of the young R\,136 cluster, contrary to what we observe.
Furthermore, with a median age of $\sim 18$\,Myr, these stars are
considerably older than the $\sim 2$\,Myr old members of R\,136, so we
should not expect any spatial relationship (either a correlation or
anti-correlation) between the distributions of the two types of objects,
unless something else related to the younger population prevents us from
detecting older PMS stars near the cluster centre.

\subsection{Erosion of circumstellar discs}

Photometric incompleteness due to the young massive stars could seem an
obvious culprit, since the younger PMS stars are on average brighter
than the older objects. However, our analysis shows that the difference
in the distribution of younger and older PMS stars remains the same as 
shown in Figure\,\ref{fig10} even when we restrict the choice of objects
to the same magnitude range (e.g. $21 < V_0 < 23$). Therefore, the
apparent paucity of older PMS stars near the centre is not caused by
selection effects in their detection.

\begin{figure}[t]
\centering
\resizebox{\hsize}{!}{\includegraphics[bb=0 0 430 410]{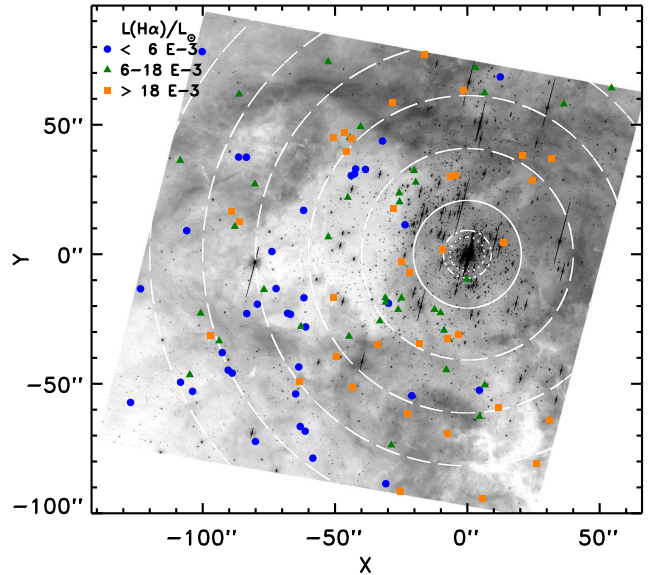}}
\caption{Spatial distribution of PMS stars with ages in the range
15--20\,Myr. Different symbols are used, according to the $L(H\alpha)$ 
luminosity of these objects, as per the legend. Stars with lower
$L(H\alpha)$ are systematically farther from the centre of R\,136.\\} 
\label{fig11}
\end{figure}

A more likely explanation is the erosion of circumstellar discs via
photo-evaporation caused by the far-ultraviolet (FUV) radiation of the
young massive members of R\,136. If the discs of PMS stars are disrupted,
the accretion process dwindles and these objects would no longer be
detectable through their H$\alpha$ excess emission. To test this
hypothesis, we have looked at how older PMS stars of various H$\alpha$
luminosity $L(H\alpha)$ are distributed in the field. In
Figure\,\ref{fig11} we show, as an example, all PMS stars with ages in
the range $15-20$\,Myr (115 objects in total) using different symbols
according to their H$\alpha$ luminosity. Dots correspond to stars with
$L(H\alpha) < 6 \times 10^{-3}$\,L$_{\odot}$, squares to $L(H\alpha) >
1.8 \times 10^{-2}$\,L$_{\odot}$, and triangles to intermediate values.
These values have been selected in such a way to have an equal number of
objects in each group. In the background we show as a reference a
negative image of the field in the H$\alpha$ band.

The distribution of the three classes of objects is not random: stars
with higher $L(H\alpha)$ are typically closer to the centre of R\,136
and located in areas where the background has a higher emission, whereas
stars with lower $L(H\alpha)$ are further away and preferentially in
regions where the background emission is low. Although systematic
differences in the mass of the three classes of objects could produce
the observed distribution of $L(H\alpha)$, this can be excluded since 
the median mass of the three groups of stars is the same at $\sim
1.3$\,\Msolar. We can also exclude that this is an artefact of our
photometry since the background is subtracted locally from each star
(see Section\,3) and therefore the diffuse H$\alpha$ nebular emission
does not affect the intrinsic H$\alpha$ brightness of individual
objects. However, a fraction of the H$\alpha$ luminosity of the stars
closest to the centre of R\,136 could in principle be due to
fluorescence of their circumstellar discs caused by the very strong
ionising radiation of the massive stars. It is, therefore, important to
quantify this effect.

According to Massey \& Hunter (1998), adopting the temperature
calibration of Vacca, Garmany \& Shull (1996), the total bolometric
luminosity of the $\sim 30$ brightest and most massive stars in R\,136
is of order $4 \times 10^7$\,L$_\odot$ and most of them are of spectral
type O4 or earlier. Based on this information and following Panagia
(1973), we have determined at which distance from the centre of R\,136
the effects of the ionising radiation impinging face-on on a
circumstellar disc of radius 100\,AU would contribute significantly to
the H$\alpha$ luminosity of its star. The median $L(H\alpha)$ value of
the PMS stars shown in Figure\,\ref{fig11} is $0.01$\,L$_\odot$, so we
have considered three cases of contamination, at approximately the
10\,\%, 50\,\% and 100\,\% level. The dotted, dot-dashed and solid
circles shown in Figure\,\ref{fig11} correspond respectively to the
distances at which the contribution to the H$\alpha$ luminosity due to
fluorescence is $ 0.01, 0.005$ and $0.001$\,L$_\odot$ (note that these
distances are necessarily upper limits since they are projected on the
plane of the sky and we have no information on the  distribution of
these objects along the line of sight). It is immediately clear that the
effects of fluorescence can  only be important for a handful of stars,
leaving the largest majority of old PMS objects in this field
unaffected. The origin of the apparent correlation between their
positions and H$\alpha$ luminosities has, therefore, to be found
elsewhere.

A natural explanation would be that stars in regions of low emission,
where the density of the gas is lower, such as those to the east of
R\,136, are much less shielded from the radiation of massive stars than
objects in denser regions, such as those to the north and west of
R\,136. Therefore, their discs have been exposed to photo-evaporation
for a longer time and their H$\alpha$ luminosity is today further
reduced. 

In support of this hypothesis, the photometry of all stars enclosed in 
the low-background region resembling a ``Christmas tree'' and centered
at ($-50\arcsec, 10\arcsec$), which is better seen in
Figure\,\ref{fig0}, is consistent with a very low extinction value,
close to the $A_V=0.22$ indicated by Fitzpatrick \& Savage (1984) as the
contribution of the Milky Way to the intervening absorption along the
line of sight. Furthermore, this region is characterised by very low
emission at IR (e.g. Indebetouw et al. 2009) and radio wavelengths (e.g.
Johansson et al. 1998; Rubio et al. 1998), confirming that it is devoid
of both dust and gas. { The region also corresponds with a
high-velocity structure seen by Chu \& Kennicutt (1994) in their
detailed H$\alpha$  echelle study of the kinematics of 30\,Dor. These
authors find a bright redshifted knot $75\arcsec$ east of R136, with a
velocity offset of $\sim 130 - 140$\,km s$^{-1}$ and very low emission.
The knot is part of a fast expanding shell to the east of the cluster that
correlates well with a  region of diffuse X-ray emission seen in images
obtained with the {\em Einstein} satellite. These authors conclude that
both the high-velocity features and the X-ray emission are related to
high-velocity shocks from fast stellar winds and supernovae. 

This picture is supported by the later study of 30\,Dor by Townsley  et
al. (2006) with the {\em Chandra X-ray Observatory}. Their observations
reveal rather strong diffuse emission in the area of the ``Christmas
tree'' (their region 6, coincident with the east fast shell of Chu \&
Kennicutt 1994) with an absorption-corrected X-ray surface brightness of
$\sim 6 \times 10^{32}$\,erg s$^{-1}$ pc$^{-2}$ in the $0.3 - 2.2$\,keV
band. A spectral fit to this emission shows a soft thermal plasma with
$k\,T=0.4$\,keV (Townsley et al. 2006). Therefore, in such an evacuated
region,} the discs of PMS stars must be much more exposed to
photo-evaporation. 

Conversely, as discussed in Section\,4, extinction towards the young
massive stars at the centre of R\,136 is both higher and highly
variable, indicating a much denser environment where gas and dust can
intercept a significant fraction of the UV radiation from the massive
central stars, thereby reducing the efficiency of the photo-evaporation
process. Indeed, most of the PMS stars with $L(H\alpha) >
0.025$\,L$_\odot$ in Figure\,\ref{fig11} are in regions of significant
$^{12}$CO$(1-0)$ emission, suggesting a high density of molecular gas,
and the column density in HI in front of R\,136 is known to be very high
($\log N=21.80 \pm 0.15 $ cm$^{-2}$; de Boer, Koornneef \& Savage 1980).

Thus, although this picture is still necessarily qualitative, it
provides a plausible explanation of the observed distribution of
H$\alpha$ luminosities of older PMS stars across the field via the
effects of photo-evaporation. These effects have been  discussed by De
Marchi et al. (2010) in the field of SN\,1987A. In that environment,
located $\sim 20\arcmin$ or 300 pc SW of 30\,Dor, the PMS stars around
the hottest massive young stars are both less numerous and fainter in
H$\alpha$ emission, suggesting that their circumstellar discs have been
eroded more efficiently by enhanced photo-evaporation. 

Unlike the relatively simple case of the SN\,1987A field where little
gas is present, the interstellar medium in 30 Dor is known to be much
denser and highly fragmented, with considerable density variations
across the field (e.g. Poglitsch et al. 1995). { From the H$\alpha$
echelle study mentioned above, Chu \& Kennicutt (1994) conclude that the
kinematics of 30\,Dor is dominated by a large number of expanding
structures, over scales of 2 -- 20\,pc and with expansion velocities of
100 -- 300\,km s$^{-1}$ and kinetic energies of $0.5 - 10 \times
10^{50}$\,erg. These structures are probably associated with
supernova remnants embedded in supershells produced by stellar winds and
supernovae explosions. 

These processes can release a large amount of kinetic energy into the
surrounding environment, and as such they could in principle contribute
to the dissipation of circumstellar discs, adding to the effects of
photo-evaporation. To better understand this, it is useful to calculate
the amount of kinetic energy possibly deposited on such a disc by a
nearby supernova explosion. We will assume hereafter a total energy
release of $E=1$\,foe $=10^{51}$\,erg per supernova explosion, a 
circumstellar radius $r=100$\,AU around a PMS star located at a 
distance of $d=10$\,pc from the supernova. The maximum fraction of
energy deposited on such a disc hit face-on by the blast can be written
as $\eta_E=E\,\pi\,r^2/d^2$ and amounts to $\sim 7.4 \times
10^{42}$\,erg for the specific parameters assumed here. The motion that
this energy release can impart onto the material of the disc depends on
the total mass and size of the disc and hence on its density, which in
turn could depend e.g. on the radial location. In the simplified case of
a homogeneous disc of mass $m_d$ that we express as a fraction $\phi$ of
the total stellar mass $m$, i.e. $m_d= \phi \, m$, the velocity $v$
imparted by the ejecta to the disc material can be expressed as: 

\begin{equation}
v= 0.86 \times \bigg[ \frac{d}{10 \, {\rm pc}} \bigg]^{-1} \, 
   \bigg[ \frac{r}{100 \, {\rm AU}} \bigg] \, 
   \left[ \phi\, \frac{m}{{\rm M}_\odot} \right]^{-1/2} \, {\rm km~s}^{-1}.
\end{equation}  

\noindent De Marchi et al. (2011) have shown that stars in the
Magellanic Clouds accrete a substantial fraction of their total mass
during the PMS phase, particularly at low masses, so for an
order-of-magnitude calculation it is not unreasonable to assume $\phi =
${\small $1/2$}. 

We can see in this way that for a star with $m=1$\,\Msolar at a distance
of 10\,pc, the supernova ejecta hitting the disc face-on would impart a
motion to its material with a velocity of $\sim 1$\,km s$^{-1}$. This
value is about a factor of 5 smaller than the escape velocity at the
edge of the disc for the typical PMS stars that we detect in the
"Christmas tree" (median mass $\sim 1.3$\,\Msolar). For a disc seen under
a smaller solid angle, the effect would be proportionally smaller. On
the other hand, less massive discs around less massive stars located
closer to the supernova would obviously experience a much stronger
effect. For instance for a star of mass {\small $1/4$}\,\Msolar at a
distance of 1\,pc from the supernova, the impulse received by the disc
material would be sufficient to exceed the escape velocity at $r \gtrsim
1$\,AU. This could significanly affect or possibly even stop the
accretion process, eventually limiting the final mass of the star. On
the other hand, with an average surface density of $\sim 0.7$\,PMS stars
per pc$^2$, only a very small fraction of the discs around objects in
this field would be severely affected by any one supernova explosion. 

In summary, it is clear that the interplay between shocks and
photo-evaporation effects due to the massive members of R\,136 can have 
important consequences on disc erosion, potentially affecting the
formation of low-mass stars in these regions. However, without more
specific kinematical information for individual objects, it remains}
difficult to quantify the expected effects on the circumstellar discs of
PMS stars using photometry alone. For a better understanding of this
complex region and a more quantitative characterisation of the
differences between the regions preferentially occupied by older PMS
stars with high and low $L(H\alpha)$, we may have to await high spectral
and spatial-resolution spectroscopic observations to obtain a map of the
amount and kinematics of the gas in the interstellar medium  along the
lines of sight towards these objects.

\subsection{Causal effects}

Finally, we briefly address the implications of our findings for the
current understanding of whether and how different generations of stars
in 30 Dor are causally connected. Walborn \& Blades (1987) and Walborn
(1991) reported three early O stars embedded in dense nebular knots to
the north and west of R\,136 and suggested that these objects could 
belong to a younger generation, triggered by the energy released by
R\,136. Using ground-based spectral classification of a large number of
OB stars in the 30 Dor nebula, Walborn \& Blades (1997) studied the ages
and spatial distribution of these objects, discovering an additional
eight early O stars embedded in dense nebular knots. From a
morphological analysis of these regions, they concluded that 30 Dor has
a two-stage starburst structure in which the energetic activity of the
$2-3$\,Myr old R\,136 has triggered the formation of a younger ($<
1$\,Myr) generation of stars at a typical projected distance of $\sim
15$\,pc, inside the dense molecular clouds that surround R\,136 to the
northeast and west. Later IR observations with NICMOS (Walborn et al.
1999; Brandner et al. 2001) resolved many of these knots into many new
IR sources, including multiple systems, clusters, and nebular
structures, indicating active star formation. 

From mainly morphological considerations in this and other Galactic star
forming regions, Walborn et al. (1999; see also Walborn 2002) suggested
that this mode of two-stage starburst, with generations separated by
$\sim 2$\,Myr, is ubiquitous. They concluded that in 30 Dor the first
generation of stars, still present, is now triggering the birth of the 
second, peripheral generation of very massive stars, which in a few
million years will be the only one left, after the most massive stars in
the core of R\,136 have disappeared. These authors predict 30 Dor to
become a giant shell H II region like N11 and NGC 604, with its most
massive stars around the periphery.

Our discovery of a considerable spread in the ages of PMS stars in this
region, with the younger generation confined to the cluster's centre and
older objects uniformly distributed across the field (see
Figure\,\ref{fig10}), raises some questions as to the validity of this
picture { at least as regards the formation of low-mass stars
triggered by older (massive) objects in their vicinities}. For instance,
if we compare the projected distribution of PMS stars with ages of less
than 2\,Myr with that of objects in the age-range $4 - 8$\,Myr we see
that they have the same centre, with the former slightly more
concentrated, rather than distributed in the periphery of the latter as
suggested in Walborn's (2002) triggered { massive} star formation
scenario. As already shown in Figure\,\ref{fig10}, the  stars around the
periphery of R\,136 are systematically older and not younger than the
members of R\,136 itself. The same conclusion had been reached by Selman
et al. (1999) from a spectroscopic study of the massive stars ($>
40$\,\Msolar) in these regions. Our findings suggest that if causal
effects are present between different generations of stars, their
interpretation cannot rely only on the relative ages and projected
distributions of the objects, but must consider proper kinematical and
possibly dynamical information. While it is undeniable that we are
seeing a case of sequential star formation, in the sense that there are
separate generations of stars over time, no clear signs of causal
effects are seen. Therefore, although the concept of triggered star
formation may well be valid, its relevance in the case of 30 Dor has yet
to be demonstrated.

\section{Summary and conclusions}

We have studied the properties of the stellar populations in the central
regions of the 30\,Dor cluster in the LMC using observations obtained 
with the WFC3 camera on board the HST in the $V$, $I$ and $H\alpha$
bands. The main results of this work can be summarised as follows.

\begin{enumerate}

\item
Our photometry delivers the deepest CMD so far obtained for this
cluster, reaching down to $V\simeq 27$ or about 3 mag deeper than
previous HST observations, and reveals for the first time a prominent
lower MS fainter than $V \simeq 22$. We attribute this feature to a
considerably older stellar population than the $\sim 3$\,Myr old stars
associated with R\,136. The CMD shows also a large number of objects in
the range $0.5 \lesssim V-I \lesssim 2$ that appear brighter than the
lower MS and that are consistent with, but extend much deeper than, the
population of PMS stars detected and studied by Sirianni et al. (2000)
and Andersen et al. (2009).

\item
The observations also reveal the presence of considerable differential
extinction across the field. We quantify the total extinction towards
massive MS stars younger than $\sim 3$\,Myr to be in the range $1.3 <
A_V < 2.1$ (respectively the 17\,\% and 83\,\% levels) with a median
value of $A_V=1.55$. We show that these stars can be used to derive a
statistical reddening correction also for stars in their vicinities, if
they have a similar age and spatial distribution. However, this is not
the case for older stars, which show a broader reddening distribution
starting already at $A_V=0.22$ (corresponding to the contribution of the
Milky Way along the line of sight towards 30\,Dor). A proper correction
for extinction, therefore, requires at least an approximate knowledge of
the age of the stars, which is not possible for all objects in the CMD.

\item
In order to distinguish relatively young objects from older stars, we 
looked for stars with H$\alpha$ excess emission, since this feature
identifies candidate PMS stars, and we studied their distribution both
across the field and in the CMD. Younger PMS stars follow the spatial
distribution of massive stars and this allows us to correct them for
reddening individually, using the extinction information derived for the
massive objects. On the other hand, older PMS objects are uniformly
distributed over the field and appear to be systematically less reddened
than younger PMS stars. We derive for them an average extinction
correction of $A_V=0.5$.

\item
Reddening correction allows us to proceed to a more accurate
identification of bona-fide PM stars. We searched for stars with
$V-H\alpha$ colour exceeding that of normal stars at the $4\,\sigma$
level or more and with $W^E_{\rm eq} > 20$\,\AA~ (or  $> 50$\,\AA\ for
stars with $T_{\rm eff} > 10\,000$\,K), in order to avoid contamination
by stars with chromospheric activity or by the winds of Be stars. A
total of 1\,159 objects satisfy these stringent conditions and we take
them as bona-fide PMS stars. 

\item 
We derived the age of PMS stars through the comparison with theoretical
PMS isochrones for metallicity $Z=0.007$, as appropriate for the LMC.
About {\small $1/3$} of these objects are younger than 4\,Myr and an
equal number are older than 12\,Myr, indicating that star formation in
this field has  proceeded over a long time, although our age resolution
does not allow us to discriminate between an extended episode and a
series of short and  frequent bursts. As regards the past 4\,Myr, we
obtain a lower limit to  the star formation rate of $\sim 200$\,\Msolar
Myr$^{-1}$ for objects in the range $\sim 0.5 - 4$\,\Msolar.

\item
We find a pronounced difference in the spatial distribution of younger
($< 4$\,Myr) and older ($> 12$\,Myr) PMS stars. The former are
concentrated near the centre of R\,136, whereas older PMS objects are
mostly located to the east of it and are remarkably absent near the
cluster centre. This effect, which is not due to photometric
incompleteness, is in general consistent with the photo-evaporation of
the circumstellar discs of older stars caused by the UV radiation of
the  massive members of R\,136. However, not all older PMS stars are
affected in the same way, since some are in regions where the high
density of the molecular gas can reduce the efficiency of the
photo-evaporation process.    

\item
Our results show that several generations of stars are present in this
field, since there are objects with ages ranging from less than 1 to
over 30\,Myr, although no clear signs of causal effects or of triggered
star formation are seen. It appears, however, that the presence of the
massive R\,136 cluster is sculpting the region and affecting low-mass
neighbouring objects through the effects of photo-evaporation.

\end{enumerate}

\vskip 1cm

\noindent
We thank an anonymous referee, whose suggestions have helped
us to improve the presentation of this work. This paper is based on
Early Release Science observations made by the WFC3 Scientific Oversight
Committee. We are grateful to the Director  of the Space Telescope
Science Institute for awarding Director's Discretionary time for this
programme. FP is grateful to the Space Science Department of ESA for
their hospitality via the visitor programme. NP acknowledges partial
support by HST--NASA grants GO--11547.06A and GO--11653.12A, and
STScI--DDRF grant D0001.82435.

\end{document}